\documentclass{elsarticle}

\usepackage{fullpage}
\usepackage{epsfig}

\newcommand{\noop}[1][1]{}
\usepackage{amsmath,amssymb,epsfig}
\usepackage{latexsym}
\usepackage{hyperref}
\usepackage{mathrsfs}
\usepackage{enumerate}

\newtheorem{mydef}{Definition}

\newtheorem{mythm}{Theorem}
\newtheorem{mylem}{Lemma}
\newtheorem{mycor}{Corollary}

\newcounter{thm} 

\newcommand{\bb}{\mathbf}
\newlength{\digitwidth}

\newcommand{\ignore}[1]{}

\begin{document}

%
%

\title{Geometry of Complex Networks and Topological Centrality}
\author{ Gyan Ranjan\corref{cor1}\fnref{fn1} }\ead{granjan@cs.umn.edu}
\author{ Zhi-Li Zhang} \ead{zhzhang@cs.umn.edu}  
\address{Dept. of Computer Science, University of Minnesota, Twin Cities, USA.}
\cortext[cor1]{Corresponding author.}
%

%
%

\begin{abstract}
We explore the geometry of complex networks in terms of an n-dimensional Euclidean 
embedding represented by the  Moore-Penrose pseudo-inverse of the graph Laplacian 
$(\bb L^+)$. 
The squared distance of a node $i$ to the origin in this n-dimensional space $(l^+_{ii})$, 
yields a topological centrality index, defined as $\mathcal{C}^{*}(i) = 1/l^+_{ii}$.   
In turn, the sum of reciprocals of individual node centralities, 
$\sum_{i}1/\mathcal{C}^*(i) = \sum_{i} l^+_{ii}$,  
or the trace of  $\bb L^+$, yields the well-known Kirchhoff index $(\mathcal{K})$, 
an overall structural descriptor for the network. 
To put in context this geometric definition of centrality, we provide alternative interpretations 
of the proposed indices that connect them to meaningful topological characteristics ---    
first, as forced detour overheads and frequency of recurrences in random walks 
that has an interesting analogy to voltage distributions in the equivalent electrical network; 
and then as the average connectedness of $i$ in all the bi-partitions of the graph.
These interpretations respectively help establish the topological centrality $(\mathcal{C}^{*}(i))$ 
of node $i$ as a measure of its overall {\em position} as well as its overall {\em connectedness} in 
the network; thus reflecting the robustness of $i$ to random multiple edge failures. 
Through empirical evaluations using synthetic and real world networks, we demonstrate how the 
topological centrality is better able to distinguish nodes in terms of their structural roles in the network 
and, along with Kirchhoff index, is appropriately sensitive to perturbations/rewirings in the network. 
\end{abstract}

%
%

\maketitle

%
%

\section{Introduction}
\label{sec:Introduction}
Unlike traditional studies on network robustness, that typically treat 
networks as combinatoric objects and rely primarily on classical 
graph-theoretic concepts (such as degree distributions, geodesics and minimum cuts), 
we explore a geometric approach as an alternative. 
To do so, we embed the network into an $n$-dimensional Euclidean space  
($n$ being the number of nodes in the network) 
represented by the Moore-Penrose pseudo-inverse of its graph Laplacian,  
denoted henceforth by $\bb L^{+}$. 
The diagonal entries of $\bb L^{+}$, denoted as $l^+_{ii}$ for the node $i$, represent 
the squared distance of node $i$ to the origin in this $n-dimensional$ space 
and provide a measure of the node's {\em topological centrality}, 
given as $\mathcal{C}^{*}(i) = 1/l^+_{ii}$.
Closer the node $i$ is to the origin in this space, or equivalently lower the $l^+_{ii}$, 
more {\em topologically central} $i$ is.       
Similarly, the trace of $\bb L^{+}$, $Tr(\bb L^{+}) = \sum_i 1/\mathcal{C}^{*}(i)$, 
determines the overall {\em volume} of the embedding and  
yields the well-known {\em Kirchhoff index} $(\mathcal{K})$, 
a structural descriptor for the network as a whole.
Once again, lower the value of $\mathcal{K}$ for a network (from amongst all possible networks 
with the same number of nodes and edges), more compact the embedding, and more 
structurally robust the overall network is. 
In short, topological centrality defines a ranking of the nodes of a given network, where as 
the Kirchhoff index provides a geometric measure to rank different networks of comparable sizes.  

In order to illustrate how the two geometric quantities defined above actually reflect the  
structural properties of the underlying network, 
and in particular to structural robustness against multiple failures,  
we provide three alternative interpretations for them in terms of: 
(a) detour overheads in random walks,   
(b) voltage distributions and the phenomenon of recurrence when the network is treated as an 
electrical circuit, and (c) the average connectedness of nodes when the network breaks into 
two, thereby making global communication untenable. We describe each of these in detail below. 

First, we show how topological centrality of a node reveals its overall {\em position} in the network. 
By equating topological centrality of a node $i$,  i.e. $\mathcal{C}^*(i) = 1/l^+_{ii}$, 
to the (reciprocal of) average detour overhead incurred when a random walk between 
any source destination pair is forced to go through $i$, we get a measure of the node's position. 
Intuitively, the average overhead incurred in such forced detours (measured in terms of the number 
of steps in the random walk) is lower if node $i$ is {\em centrally} positioned in the network 
(higher  $\mathcal{C}^*(i)$ and lower $l^+_{ii}$) and higher if $i$ is {\em peripheral}. 
Secondly, we show how $\mathcal{C}^*(i)$ captures voltage distribution when 
the network is transformed into an equivalent electrical network (EEN). This, in turn, 
is related to the probability with which a random detour through $i$ returns to the source node; 
referred to as the phenomenon of {\em recurrence} in random walk literature. 
To be precise, higher $\mathcal{C}^*(i)$ implies that a random detour through node $i$ forces 
the random walk between any source destination pair to return to the source node with lower probability, 
thereby incurring lower detour overhead. Both of these interpretations, namely average 
detour overhead and probability of recurrence, therefore, demonstrate how $\mathcal{C}^*(i)$ quantifies 
the overall position of node $i$ in the network.  
Finally, we establish how $\mathcal{C}^*(i)$ captures the overall connectedness 
of node $i$. To do so, we equate it to the number of nodes that $i$ can communicate with when 
a sub-set of edges in the network fail in such a way that the network is partitioned into two connected 
sub-networks. As connected bi-partitions represent a regressed state of the network when 
not all pairs of nodes can maintain communication, 
a higher value of $\mathcal{C}^*(i)$, implies that $i$ is present in the larger of the two 
sub-networks, on an average, in such bi-partitions. Thus, $\mathcal{C}^*(i)$ reflects 
the immunity/vulnerability of node $i$ towards multiple edge failures in the network, 
a distinct topological characteristic. 

Through numerical simulations using synthetic and realistic network topologies, 
we demonstrate that our new indices better characterize robustness of nodes in 
network, both in terms of position as well as connectedness, as compared to other existing metrics 
(e.g. node centrality measured based on degree, shortest paths, etc.). 
A rank-order of nodes by their topological centralities $(\mathcal{C}^*(i))$ helps distinguish them 
in terms of their structural roles (such as core, gateway, etc.). Also, topological centrality and  
Kirchhoff index, are both appropriately sensitive to local perturbations in the network, a 
desirable property not displayed by some of the other popular centrality indices 
in literature (as shown later in this paper). 

The rest of the paper is organized as follows:   
we begin by providing a brief overview of several structural indices, characterizing node centrality 
as well as overall descriptors for networks, found in literature in \textsection\ref{sec:RelWork}.
\textsection\ref{sec:Geometry} introduces  a geometric embedding of  the network using the eigen-space 
of $\bb L^+$, topological centrality and Kirchhoff index as measures of robustness.
\textsection\ref{sec:StructCentAndRandomWalks} demonstrates how  topological centrality of a 
node reflects the average detour overhead in random walks through a particular node in question followed 
by its equivalence to the probability of recurrence.
In \textsection\ref{sec:Partitions} we show how topological centrality captures the average connectedness of  nodes in the bi-partitions of a network. 
\textsection\ref{sec:Empirical} presents comparative empirical analysis with simulated as well as real world 
networks while in \textsection\ref{sec:Complexity} we analyze the computational complexity of the proposed 
metrics with respect to others popular in literature. 
Finally, the paper is concluded with a discussion of future work in \textsection\ref{sec:Conclusion}.
\

%
%

%
%

\section{Related Work}
\label{sec:RelWork}
Robustness of  nodes to failures in complex networks is dependent on their 
overall {\em position} and {\em connectedness} in the network.
Several centralities, that characterize position and/or connectedness of nodes 
in complex networks in different ways, have therefore been proposed in literature.
Perhaps the simplest of all is degree --- the number of edges incident on a node. 
Degree is essentially a {\em local} measure i.e. a first oder/one-hop connectedness index. 
A second-order variant called {\em joint-degree}, 
given by the product of degrees of a pair of nodes that are connected by an edge in the 
network, is also in vogue. However, except in {\em scale free} networks that display the 
so called {\em rich club connectivity} \cite{Barabasi00, Faloutsos99, Barabasi01}, 
neither degree nor joint-degree determine the overall position or the connectedness of nodes.

A class of structural indices called {\em betweennesses}, namely shortest path/geodesic $(GB)$
\cite{Freeman77, Freeman79}, flow $(FB)$ \cite{Freeman91} and random-walk $(RB)$ 
\cite{Newman05} betweenness respectively quantify the positions of nodes, with 
respect to source destination pairs in the network. The set of betweennesses, therefore, 
reflect the role played by a node in the communication between other node-pairs in the network 
and are not the measures of a node's own connectedness. 

Another popular centrality measure is geodesic closeness $(GC)$\cite{Freeman77, Freeman79}. 
It is defined as the (reciprocal of) average shortest-path distance of a node from all other nodes 
in the network. Clearly, geodesic closeness is a $p^{th}$-order measure of connectedness 
where $p = \{1, 2, ..., \delta\}$, $\delta$ being the geodesic diameter of the graph, and is better 
suited for characterizing global connectedness properties than the aforementioned indices.
However, communication in networks  is not always confined to shortest paths alone and $GC$ 
being geodesic based, ignores other alternative paths between nodes, however competitive they 
might be, and thus only partially captures connectedness of nodes. Some all paths based 
counterparts of geodesic closeness include information centrality \cite{Stephenson89} and 
random-walk centrality \cite{Noh04}, that use random-walk based approach to measure centrality.   
In \cite{Borgatti05}, several centrality measures based on network flow, and collectively referred 
to as {\em structural centrality}, have also been discussed in great detail.  

Recently, subgraph centrality $(SC)$ --- the number of subgraphs of a graph that a node 
participates in --- has also been proposed \cite{Estrada05}. In principle, a node with high 
subgraph centrality, should be better connected to other nodes in the network through 
redundant paths. 
Alas, subgraph centrality is computationally intractable and the proposed 
index in \cite{Estrada05} approximates subgraph centrality by the sum of lengths of all 
{\em closed} walks, weighed in inverse proportions by the factorial of their lengths.
This inevitably results in greater correlation with node degrees as each edge contributes 
to closed random-walks of lengths $2, 4, 6, ...$  and thus introduces local connectivity bias. 
In a subsequent paper, Estrada et al introduce the concept of vibrations to measure node 
vulnerability in complex networks \cite{Estrada10}. Their index called {\em node displacement} 
bears significant resemblance to information centrality and has interesting analogies to physical 
systems. However, once again the true topological significance of the centrality measure, 
in terms of connectedness is wanting. 

Our aim in this work, therefore, is to provide an index for robustness of 
nodes in complex networks,  that effectively reflects both the position and connectedness 
properties of nodes and consequently, by extension, of the overall network. 
\

%
%

%
%

\section{Geometric Embedding of Networks using $\bb L^+$, Topological Centrality and Kirchhoff Index}
\label{sec:Geometry}
In studying the {\em geometry} of networks, we first need to embed a network, 
represented abstractly as a graph, into an appropriate space endowed with a 
metric function (mathematically, a metric space).     
In this section, we first describe just such a metric space in terms of the the 
Moore-Penrose pseudo-inverse of the combinatorial Laplacian (\textsection\ref{subSec:Embed}). 
Next, using the geometric attributes of this metric space, we define the topological centralities for 
individual nodes as well as the Kirchhoff index for the network as a whole 
(\textsection\ref{subSec:TopCentKirchhoff}).   
\subsection{Network as a Graph, the Laplacian and a Euclidean Embedding}
\label{subSec:Embed}
Given a complex network, we represent its topology as a weighted undirected   
graph, $G = (V, E, W)$, where $V(G)$ is the set of nodes; 
$E(G)$ the set of edges; and 
$W = \{w_{uv} \in \Re^{+} : e_{uv} \in E(G)\}$ is a set of weights assigned to each 
edge of the graph (here $\Re^{+}$ denotes the set of nonnegative real numbers). 
These weights can be used to represent a variety of affinity/distance measures, 
depending upon the context, such as latency or capacity in traditional communication networks, 
as well as friendship and acquaintance type relationships in the social counterparts. 
We define $\bb A = [a_{ij}]$ as the affinity matrix of $G(V, E, W)$, 
such that $a_{ij} = w_{ij}$ represents the affinity between nodes $i$ and $j$: 
larger the value of $a_{ij}$ is,  {\em closer} the nodes $i$ and $j$ are. 
For a simple graph where $w_{ij} \in \{0, 1\}$, $\bb A$ is simply the standard adjacency 
matrix of the graph $G$. 
Let $n = |V(G)|$ be the number of nodes in $G$ (also called the {\em order} of G). 
For $1\leq i \leq n$, we define $d(i) = \sum_{j} a_{ij}$, which is the (generalized) degree 
of node $i$. The sum of node degrees is often referred to as the {\em volume} of the graph, 
given as $Vol(G) = \sum_{i \in V(G)} d(i)$. It is easy to see that when the graph is unweighted, 
$Vol(G) = 2 |E(G)|$. 

The {\em combinatorial Laplacian} of the graph $G(V, E, W)$,   
is defined as $\bb L = \bb D - \bb A$, where $\bb D = [d_{ii}] = d(i)$ is a diagonal matrix with 
the node degrees on the diagonal. 
$\bb L$ is a square, symmetric, doubly-centered (all rows and columns sum up to 0) and 
positive semidefinite matrix \cite{Ben-IsraelGreville03} with $n$ non-negative eigen values. 
By convention, we order the eigen values of $\bb L$ in decreasing order of 
magnitude as $[\lambda_1 \geq \lambda_2 \geq ... \lambda_{n-1} > \lambda_n = 0]$.  
Similarly for $1 \leq i \leq n$, let $\bb u_{i}$ be the corresponding eigenvector of $\lambda_{i}$. 
If the network is connected (which we assume to be the case henceforth), the smallest eigen 
value $\lambda_n = 0$ and its corresponding eigen vector $\bb u_n = [1, 1, ..., 1]$ are both unique.    
Also, the eigen vectors of $\bb L$ are mutually perpendicular, i.e. $\bb u_i \cdot \bb u_j = 0, \forall i, j$ 
(where $(\cdot)$ is the inner product). 
Therefore, the matrix of eigen vectors $\bb U = [\bb u_1, \bb u_2, ... \bb u_n]$ represents an 
orthonormal basis for an $n$-dimensional Euclidean space. 
In short, we say that the Laplacian $\bb L$ admits an eigen decomposition of the form 
$\bb L = \bb U \Lambda \bb U'$, where $\Lambda$ is the diagonal matrix 
$\Lambda = [\lambda_{ii}] = \lambda_i$ and $\bb U$ is set of $n$ orthonormal eigen vectors.   

Like $\bb L$, its Moore-Penrose pseudo-inverse $\bb L^+$ is also square, symmetric, doubly-centered 
and  positive semi-definite \cite{Fouss07}. It thus admits an eigen decomposition of the form, 
$\bb L^+ = \bb U \Lambda^{+} \bb U'$, where $\bb \Lambda^{+}$ is a diagonal matrix consisting of 
$\lambda^{-1}$ if $\lambda_i > 0$, and 0 if $\lambda_i = 0$. 
It is the eigen space of $\bb L^+$ (derived from the eigen space of $\bb L$) that is of interest to us. 
Let $\bb X = \Lambda^{+1/2} \bb U'$. We can therefore rewrite $\bb L^+$ as: 
\begin{equation} 
\label{equ:LpEmbed}
\bb L^{+} =   \bb U \Lambda^{+} \bb U' = \bb X' \bb X  
\end{equation}
The form in $(\ref{equ:LpEmbed})$ above, together with the fact that the matrix $\bb U$ is an orthonormal basis 
for $\Re^n$, implies that the matrix $\bb X$ represents an embedding of the network in an $n$-dimensional 
Euclidean space  (for details please refer \cite{Fouss07} and the references therein).  

Each node $i \in V(G)$ of the network is now represented in terms of a point in this $n$-dimensional space, 
characterized by the position vector $\bb x_i$, i.e. the $i^{th}$ column of $\bb X$. 
Also, as $\bb L^+$ is doubly-centered (all rows and columns sum to 0), 
the centroid of the position vectors for the set of nodes lies at the origin of this n-dimensional space. 
Thus, the squared distance of node $i$ from the origin (or the squared length of the position vector) 
is exactly the corresponding diagonal entry of $\bb L^+$ i.e. $||\bb x_i||^2_2 = l^+_{ii}$. 

Similarly, the squared distance between two nodes $i, j \in V(G)$, is given by 
 $||\bb x_i - \bb x_j||^2_2 = l^+_{ii} + l^+_{jj} - l^+_{ij} - l^+_{ji}$. 
This pairwise distance is also called the {\em effective resistance} distance \cite{KleinRandic93}, 
which in turn is a scaled version of the expected length of a random commute between nodes 
$i$ and $j$ in the underlying graph (details in subsequent sections).    
\subsection{Topological Centrality and Kirchoff Index} 
\label{subSec:TopCentKirchhoff}
Based on the geometric embedding of the graph using $\bb L^+$ described above, we now put 
forth two metrics. First, a rank order for individual nodes in terms of their relative 
robustness characteristics called {\em topological centrality}, defined as: 
\begin{mydef}
Topological centrality of node $i \in V(G)$: 
\begin{equation}
\label{equ:TopCentLpDiag}
\mathcal{C}^*(i) = 1/l^+_{ii}, ~~~ \forall i \in V(G) 
\end{equation}
\end{mydef}
Thus, closer a node $i$ is to the origin in this n-dimensional space, i.e. lower the numerical value of $l^+_{ii}$, 
more topologically central it is, i.e. higher $\mathcal{C}^*(i)$. More importantly, and as we shall demonstrate 
in the sections to follow, higher the topological centrality of a node, more {\em centrally} located it is in the 
network (structurally) and greater its robustness to multiple edge failures in the network.   
But before we proceed, a brief discussion of the definition in $(\ref{equ:TopCentLpDiag})$ 
is warranted to put the topological centrality metric in context. 
The element $l^+_{ii}$ in $\bb L^+$ can be rewritten in terms of the elements Laplacian spectrum, 
as follows: 
\begin{equation}
\label{equ:TopCentLpDiagSpec}
l^+_{ii} = \sum_{j=1}^{n-1} \frac{\bb u_{ji}^2}{\lambda_j}   
\end{equation}
Thus, the topological centrality of a node is a function of the entire eigen spectrum of the 
graph Laplacian ($\bb L$). Clearly, the contribution made to the overall value of $l^+_{ii}$ 
by a particular eigen pair $(\lambda_i, \bb u_i)$ is determined by the ratio $\bb u_{ji}^2/{\lambda_j}$. 
This attribute, though simple essence, is rather important and sets the topological centrality measure 
apart. 
The use of matrix spectra to study structural properties of networks is quite popular in literature. In  
\cite{Dorogovtsev03} the spectra of the adjacency and related ensemble matrices have been studied, 
whereas in \cite{Mitrovic09} a subset of leading eigen vectors of the graph Laplacian 
and its normalized counterparts have been used for {\em localizing} a subset of closely 
connected nodes (or communities). 
Our topological centrality $(\mathcal{C}^*(i))$, seen in this light, is a more generalized metric  
that extends previously known localization approaches (similar to that in  \cite{Mitrovic09}) to the granularity 
of individual nodes. 

Next, we define a structural descriptor for the overall robustness of the network called {\em Kirchhoff index}, 
as:
\begin{mydef}
Kirchhoff index for $G(V, E)$: 
\begin{equation}
\label{equ:KirchhoffIndex}
\mathcal{K}(G) = Tr(\bb L^+) = \sum_{i=1}^n l^+_{ii} = \sum_{i=1}^n 1/ \mathcal{C}^*(i)
\end{equation}
\end{mydef}
Geometrically, more compact the embedding is, lower the value of $\mathcal{K}(G)$ and 
more robust the network $G$ is (shown in latter sections) 
\footnote{In literature, and in particular in \cite{Xiao03} (c.f. corollary 2.3), 
the Kirchhoff index for a network sometimes appears as 
$\mathcal{K}(G)  = n ~Tr(\bb L^+)$, i.e. with a scaling factor of $n$ over what 
we have defined above in \ref{equ:KirchhoffIndex}. 
However, in this work, our aim is to use $\mathcal{K}(G)$ as a comparative measure of 
robustness for two networks of the same order (i.e. same values of $n$) and volume 
(as described in subsequent sections). 
Therefore, we do away with the constant $n$ from  the definition of $\mathcal{K}(G)$ 
for the rest of this work.}. 
Kirchhoff index has been widely used  to 
model molecular strengths in the mathematical chemistry literature 
\cite{Palacios01b, Palacios10a, Palacios01a, Palacios10b, Xiao03, Zhou08} as well as 
in linear algebra \cite{Bendito10}.  
However, its true topological significance has scarcely been explored and/or demonstrated. 
Note,  
\begin{equation}
\mathcal{K}(G) = Tr(\bb L^+) = \sum_{i=1}^{n-1} \frac{1}{\lambda_i}
\end{equation}
\noindent Once again, we see that the global structural descriptor is a function of the overall 
Laplacian spectrum. Therefore, Kirchhoff index can be thought of as a generalized analogue 
of the much celebrated {\em algebraic connectivity} of the graph \cite{Fiedler73, Fiedler75}, 
which is measured in terms of the second smallest eigen-value of the Laplacian i.e. $\lambda_{n-1}$, 
or, equivalently, the largest eigen value of $\bb L^+$. 

In what follows, we demonstrate how these two metrics indeed reflect robustness of nodes and the 
overall graph respectively, first through rigorous mathematical analysis resulting in closed form 
representations and  then with empirical evaluations over realistic network topologies.
\

%
%

%
%

\section{Topological Centrality, Random Walks and Electrical Voltages}
\label{sec:StructCentAndRandomWalks}
To show that topological centrality $(\mathcal{C}^*(i))$ indeed captures the overall position of a node, 
we relate it to the lengths of random-walks on the graph.
In \textsection\ref{subsec:Detour}, we demonstrate how $\mathcal{C}^*(i)$ is related to the average  
overhead incurred in random {\em detours} through node $i$ as a {\em transit} vertex. 
Next in \textsection\ref{subsec:RecurVoltAndElectNet}, we provide an electrical interpretation for it in terms 
of voltages and the probability with which a random detour through node $i$ returns to the source node. 
\subsection{Detours in Random Walks}
\label{subsec:Detour}
A simple random walk $(i \rightarrow j)$, is a discrete stochastic process 
that starts at a node $i$, the source, visits other nodes in the graph $G$ 
and stops on reaching the destination $j$  \cite{GobelJagers74}. In contrast, 
we define a {\em random detour} as:
\begin{mydef}
Random Detour $(i \rightarrow k \rightarrow j)$: A random walk starting from 
a source node $i$, that must visit a transit node $k$, before it reaches the 
destination $j$ and stops.
\end{mydef}
Effectively, such a random detour is a combination of two simple random walks: 
$(i \rightarrow k)$ followed by $(k \rightarrow j)$. We quantify the difference 
between the random detour $(i \rightarrow k \rightarrow j)$ and the simple random 
walk $(i \rightarrow j)$ in terms of the number of steps required to complete 
each of the two processes given by hitting time.
\begin{mydef}
Hitting Time $(H_{ij})$: The expected number of steps in a 
random walk starting at node $i$ before it reaches node $j$ for the first 
time. 
\end{mydef}
Clearly, $H_{ik} + H_{kj}$ is the expected number of steps in the random 
detour $(i \rightarrow k \rightarrow j)$. Therefore, the overhead incurred is: 
\begin{equation}
\Delta H^{i \rightarrow k \rightarrow j}  = H_{ik} + H_{kj} - H_{ij}
\label{equ:DetOverhead}
\end{equation}
Intuitively, more peripheral transit $k$ is, greater the overhead in (\ref{equ:DetOverhead}). 
The overall peripherality of $k$ is captured by the following average:
\begin{equation}
\Delta H^{(k)}
= \frac{1}{n^2 ~Vol(G)} 
\sum_{i = 1}^{n} \sum_{j = 1}^{n} \Delta H^{i \rightarrow k \rightarrow j}
\label{equ:RandEcc}
\end{equation}
Alas, hitting time is not a Euclidean distance as $H_{ij} \neq H_{ji}$ 
in general. An alternative is to use commute time 
$C_{ij} = H_{ij} + H_{ji} = C_{ji}$, a metric, instead. More importantly \cite{KleinRandic93}, 
\begin{equation}
C_{ij} = Vol(G) (l^+_{ii} + l^+_{jj} - l^+_{ij} - l^+_{ji})
\end{equation}
and in the overhead form $(\ref{equ:DetOverhead})$, (non-metric) hitting and 
(metric) commute times are in fact equivalent 
(see propositions $9-58$ in \cite{Kemeny66}):
\begin{equation}
\Delta H^{i \rightarrow k \rightarrow j} = (C_{ik} + C_{kj} - C_{ij})/2 
= \Delta H^{j \rightarrow k \rightarrow i}
\end{equation}
We now exploit this equivalence to equate the cumulative detour overhead 
through transit $k$ from (\ref{equ:RandEcc}) to $l^+_{kk}$ in the following 
theorem.\newline
%
%
%
%
\noindent\begin{mythm} \label{thm:RandEccLp}
\begin{equation}
\Delta H^{(k)} = l^+_{kk}
\end{equation}
\end{mythm}
\textbf{Proof:}
Using $\Delta H^{i \rightarrow k \rightarrow j} = (C_{ik} + C_{kj} - C_{ij})/2$:
\begin{eqnarray*}
\Delta H^{(k)} &=& 
\frac{1}{2 n^2  ~Vol(G)} 
\sum_{i = 1}^{n} \sum_{j = 1}^{n} C_{ik} + C_{kj} - C_{ij}
\end{eqnarray*}
Observing $C_{xy}= Vol(G)~(l^{+}_{xx} + l^{+}_{yy}- 2 l^{+}_{xy})$ 
\cite{KleinRandic93} and that $\bb L^+$ is doubly centered 
(all rows and columns sum to $0$) \cite{Fouss07}, we obtain the proof.

\noindent $\square$

\noindent Therefore, a low value of $\Delta H^{(k)}$ implies higher $\mathcal{C}^*(k)$ and more 
structurally central node $k$ is in the network. Theorem \ref{thm:RandEccLp} is interesting for 
several reasons. First and foremost, note that:
\begin{equation}
\sum_{j=1}^{n} C_{kj} = Vol(G) ~(n~ l^{+}_{kk} + Tr(\bb L^+))
\end{equation}
As $Tr(\bb L^+)$ is a constant for a given graph and an invariant with 
respect to the set $V(G)$, we obtain 
$l^+_{kk} \propto \sum_{j=1}^{n} C_{kj}$;  
lower $l^+_{kk}$ or equivalently higher $\mathcal{C}^*(k)$, implies shorter 
average commute times between $k$ and the rest of the nodes in the graph 
on an average. 
Moreover,
\begin{equation}
\mathcal{K}(G) = Tr(\bb L^+)=\sum_{k = 1}^n l^+_{kk} = \frac{1}{2n Vol(G)} \sum_{k=1}^{n}\sum_{j=1}^{n} C_{kj}
\end{equation} 
As $\mathcal{K}(G)$ reflects the average commute time between any pair of nodes in the network, 
it is a measure of overall connectedness in $G$. For two networks of the same order $(n)$ and 
volume $(Vol(G))$, the one with lower $\mathcal{K}(G)$ is better connected on an average. 
\begin{figure}[t]
\centerline{\begin{tabular}{c}
\scalebox{0.8}{\includegraphics[width = 100mm]{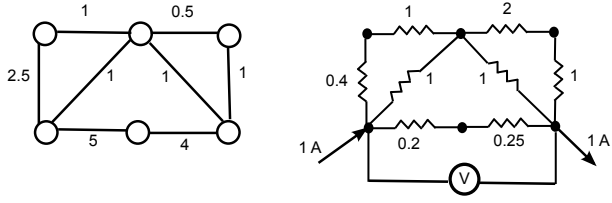}}
\end{tabular}}
\caption{A simple graph $G$  and its EEN.}
\label{fig:GraphEEN}
\end{figure}
\subsection{Recurrence, Voltage and Electrical Networks}
\label{subsec:RecurVoltAndElectNet}
Interestingly, the detour overhead in  $(\ref{equ:DetOverhead})$ is related 
to {\em recurrence} in random walks --- the expected number of times a 
random walk  $(i \rightarrow j)$ returns to the source $i$  \cite{DoyleSnell84}. 
We now explore how recurrence in detours related to topological centrality 
of nodes. But first we need to introduce some terminology.

The equivalent electrical network (EEN)  \cite{DoyleSnell84} for $G(V, E, W)$ 
is formed by replacing an edge $e_{ij} \in E(G)$ with a {\em resistor}. 
The resistance of this resistor is equal to $w_{ij}^{-1}$ (see Fig. \ref{fig:GraphEEN}), 
where $w_{ij}$ is the affinity between nodes $i$ and $j$, or equivalently, the weight 
associated with the edge $e_{ij}$ in the graph $G$. 
The {\em effective resistance} ($\Omega_{ij}$) is defined as the voltage developed 
across a pair of terminals $i$ and $j$ when a unit current is injected at $i$ and 
is extracted from $j$, or vice versa.
In the EEN, let $V^{ij}_{k}$ be the voltage of node $k$ when a unit current is 
injected at $i$ and a unit current is extracted from $j$. From \cite{Tetali91}, 
$U^{ij}_{k} = d(k) V^{ij}_{k}$; where $U^{ij}_{k}$ is the expected number of times a 
random walk $(i \rightarrow j)$ visits node $k$. 
Substituting $k=i$ we get, $U^{ij}_i = d(i) V^{ij}_i$; the expected number of 
times a random walk  $(i \rightarrow j)$ returns to the source $i$. For a finite 
graph $G$, $U^{ij}_i > 0$.
The following theorem connects recurrence to the 
detour overhead. 
\begin{mythm} \label{thm:DetOverheadRetCount}
\begin{eqnarray*}
\Delta H^{i \rightarrow k \rightarrow j} 
&=& \frac{Vol(G) ~ (U^{ik}_{i} + U^{kj}_{i} - U^{ij}_{i})}{d(i)}
\\
&=&
Vol(G) ~ (V^{ik}_{i} + V^{kj}_{i} - V^{ij}_{i})
\end{eqnarray*}
\end{mythm}
\textbf{Proof:} From \cite{Tetali91} we have, 
$\Delta H^{i \rightarrow k \rightarrow j} = d(i)^{-1} ~Vol(G) ~ U^{jk}_i$. 
The rest of this proof follows by proving 
$U^{jk}_{i} = U^{ik}_{i} + U^{kj}_{i} - U^{ij}_{i}$.

From the \textit{superposition principle} of electrical current, 
we have $V^{xz}_{x} = V^{xz}_{y} + V^{zx}_{y}$. Therefore,
\begin{eqnarray*}
V^{ik}_{i} + V^{kj}_{i} - V^{ij}_{i}   
&=& 
V^{ik}_{j} + V^{ki}_{j} +  V^{kj}_{i} - V^{ij}_{k} + V^{ji}_{k}
\\
&=&
V^{ik}_{j} + (V^{ki}_{j} +  V^{kj}_{i} - V^{ij}_{k} - V^{ji}_{k}) 
\end{eqnarray*}
From the \textit{reciprocity principle}, $V^{xy}_{z} = V^{zy}_{x}$. Therefore,
$V^{ik}_{i} + V^{kj}_{i} - V^{ij}_{i}  = V^{jk}_{i}$.
Multiplying by $d(i)$ on both sides we obtain the proof.

\noindent $\square$

The term $(U^{ik}_{i} + U^{kj}_{i}) - U^{ij}_{i}$ can be interpreted as the expected 
extra number of times a random walk returns to the source $i$ in the random 
detour $(i \rightarrow k \rightarrow j)$ as compared to the simple random 
walk $(i \rightarrow j)$. 
Each instance of the random process that returns to the source, 
must effectively start all over again. Therefore, more often the walk returns 
to the source greater the expected number of steps required to complete the 
process and less central the transit $k$ is, with respect to the source-destination 
pair $(i, j)$.

Therefore, $\Delta H^{(k)}$, that is the average of $\Delta H^{i \rightarrow k \rightarrow j}$ 
over all source destination pairs, tells us the average increase in recurrence caused 
by node $k$ in random detours between any source destination pair in the network. 
Higher the increase in recurrence, i.e. $\Delta H^{(k)}$, lower the magnitude of 
$\mathcal{C}^*(k)$ and less structurally central the node $k$ is in the network. 
\

%
%

%
%

\section{Connected Bi-Partitions of a Network}
\label{sec:Partitions}
\begin{figure}
\centerline{\begin{tabular}{cc}
\scalebox{0.75}{\includegraphics[width = 40mm]{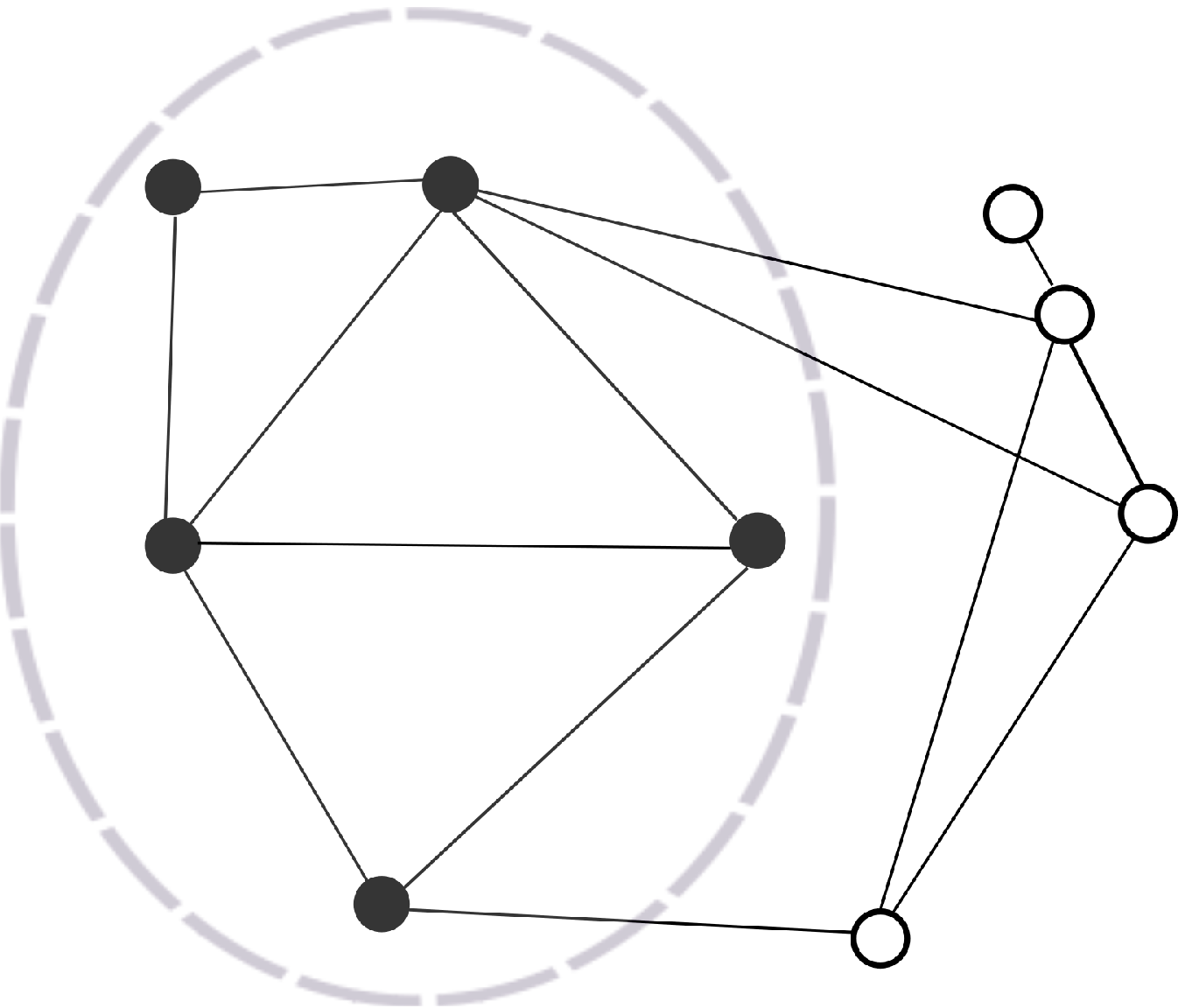}}
&
\scalebox{0.75}{\includegraphics[width = 40mm]{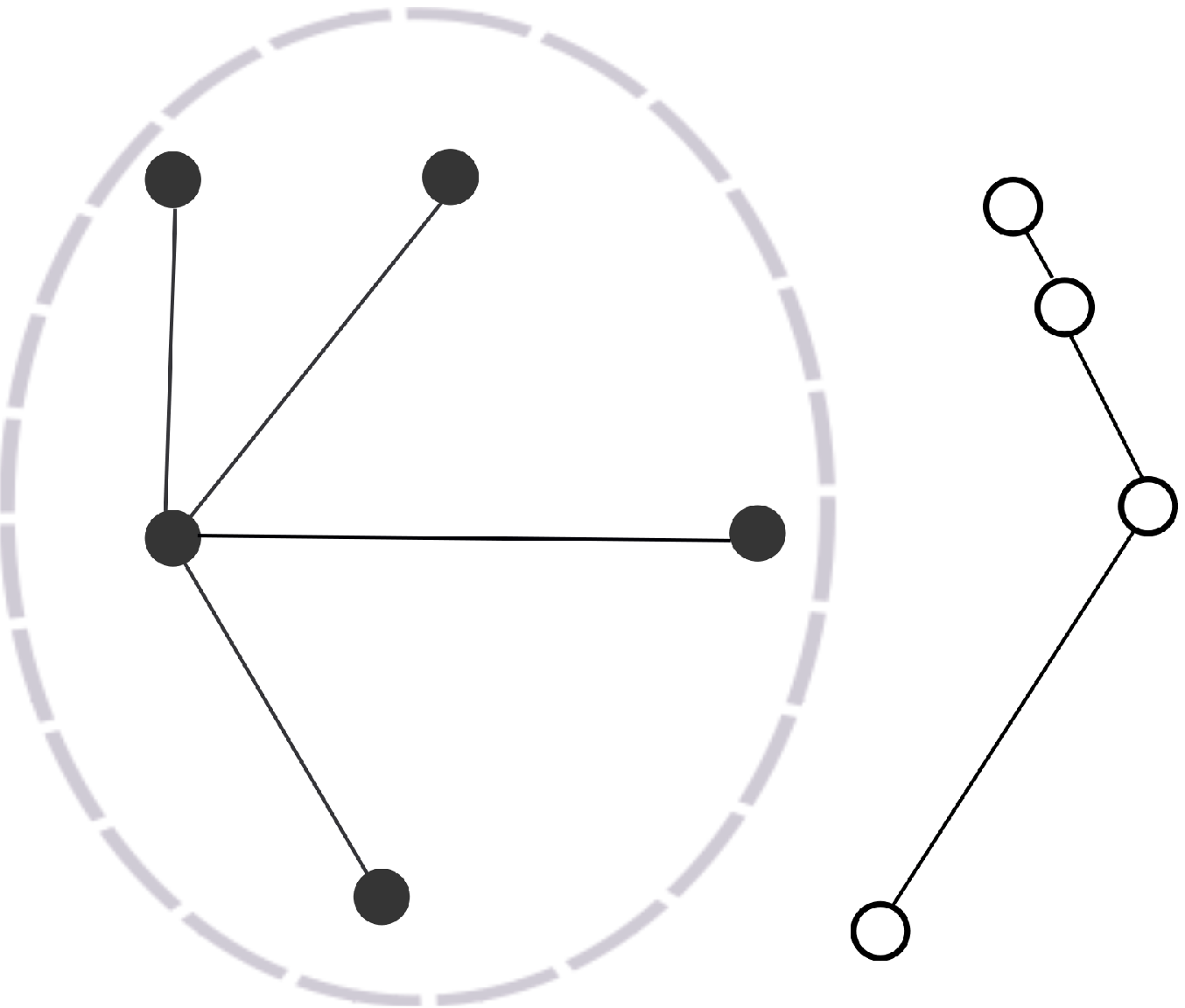}}
\\
(a) Partition $P=(S,S')$ & (b) Spn. forest in $P=(S,S')$
\\
\end{tabular}}
\caption{Partitions and spanning forests of a graph.}
\label{fig:CutPartition}
\end{figure}
Having thus far established that nodes with higher $\mathcal{C}^*(i)$, 
are more structurally central in the network, we now turn to average connectedness 
of nodes in this section. To show how structural centralities of nodes capture their 
immunity/vulnerability to random failures in the network, we study their connectedness 
in all the bi-partitions of a graph.
\subsection{Connected Bi-partitions}
\begin{mydef}
Bi-partition $(P=(S, S'))$: A cut of the graph $G$ which contains exactly two 
mutually exclusive and exhaustive connected subgraphs $S$ and $S'$.
\end{mydef}
Let, $V(S)$ and $V(S')$ be the mutually exclusive and exhaustive subsets of 
$V(G)$, $E(S)$ and $E(S')$, the sets of edges in the respective components 
$S$ and $S'$ of $P$ and $E(S, S')$, the set of edges that violate $P$ i.e. have 
one end in $S$ and the other in $S'$.
Also, let $\mathcal{T}(S)$ and $\mathcal{T}(S')$ be the set of spanning trees 
in the respective component sets $S$ and $S'$.
We denote by $\mathcal{P}(G)$, the set of all bi-partitions of $G(V, E)$. 
Clearly, a given $P=(S, S')$ represents a state of the network in which $E(S, S')$ 
have failed. A node $i \in V(S)$ stays connected to $|V(S)| - 1$ nodes 
and gets disconnected from $|V(S')|$ nodes.
In the following relationship we show how topological centrality of node $i$ is related to a  
weighted sum of $|V(S)'|$ over all the bi-partitions $P \in \mathcal{P}(G)$ of the network.
\begin{mythm}
\label{thm:LpPropPartitions}
\begin{equation}
l^+_{ii} \propto \sum_{P \in \mathcal{P}(G)}^{i \in V(S)} |\mathcal{T}(S)| |\mathcal{T}(S')| |V(S')|
\end{equation}
\end{mythm}
\textbf{Proof sketch:} The following result, due to Chebotarev et al.  \cite{ChebShamis97, ChebShamis98b}, 
forms the basis of our proof. Let $\mathcal{F}_{x}$ be the set of spanning rooted forests of $G(V, E)$ with 
$x$ edges. Precisely, $F_{x} \in \mathcal{F}_{x}$, is a spanning acyclic subgraph of $G$ with the same node 
set as $G$ and is composed of exactly $n - x$ trees with one node marked as a {\em root} in each of 
these $x$ trees. Let $\mathcal{F}^{ii}_{x}$, be the subset of $\mathcal{F}_{x}$ in which node $i$ is the root 
of the tree in which it belongs. Then,
\begin{equation}
\label{equ:DenseSRFs}
l^+_{ii} 
= \frac{\varepsilon(\mathcal{F}^{ii}_{n-2}) - \frac{1}{n} \varepsilon(\mathcal{F}_{n-2})}
{\varepsilon(\mathcal{F}_{n-1})}
\end{equation}
Here, $\varepsilon(\cdot)$ simply represents the cardinality of the input set 
(see  \cite{ChebShamis97, ChebShamis98b} for details). It is easy to see that 
$\varepsilon(\mathcal{F}_{n-1}) = n |T(G)|$, as a spanning forest with $n - 1$ edges is 
 a spanning tree, and each spanning tree has exactly $n$ possible choices of roots. 
Also,  $\varepsilon(\mathcal{F}_{n-2})$ and  $\varepsilon(\mathcal{F}_{n-1})$ 
are invariants over the set of vertices $V(G)$ for a given graph.
Hence, $l^+_{ii} \propto \varepsilon(\mathcal{F}^{ii}_{n-2})$.
The rest of the proof follows from the results of the following lemmas:
\begin{mylem} 
\label{lem:RandEccPerPartition}
Let $\mathcal{F}^{ii}_{n - 2|P}$ be the set of spanning forests with 
$n - 2$ edges (or exactly two trees) rooted at node $i$ in a given bi-partition $P=(S, S')$ and 
$\mathcal{T}(S)$, $\mathcal{T}(S')$ be the set of spanning trees in $S$ and 
$S'$ respectively. If $i \in V(S)$ then,
$$\varepsilon(\mathcal{F}^{ii}_{n - 2|P}) = |\mathcal{T}(S)| ~ |\mathcal{T}(S')| ~ |V(S')|$$ 
\end{mylem}
\textbf{Proof :} Let $T_1 \in \mathcal{T}(S)$ and $T_2 \in \mathcal{T}(S')$. 
Clearly, $|E(T_1)| = |V(S)| - 1$ and $|E(T_2)| = |V(S')| - 1$. 
As $|V(S)| + |V(S')| = |V(G)|$ and $|E(T_1) \cap E(T_2)| = 0$, 
$|E(T_1) \cup E(T_2)| = |V(S)| - 1+ |V(S')| - 1 = n - 2$.
Each such pair $(T_1, T_2)$ is a spanning forest of $n - 2$ edges. 
Given $i$ is the root of $T_1$ in $S$, we can choose $|V(S')|$ roots for 
$T_2$ in $S'$. There being $|\mathcal{T}(S)| ~ |\mathcal{T}(S')|$ 
such pairs:
$\varepsilon(\mathcal{F}^{ii}_{n - 2|P}) = |\mathcal{T}(S)| ~ |\mathcal{T}(S')| ~ |V(S')|$.

\noindent $\square$

\noindent By symmetry, for $j \in V(S'):$
$$\varepsilon(\mathcal{F}^{jj}_{n - 2|P}) = |\mathcal{T}(S)| ~ |\mathcal{T}(S')| ~ |V(S)|$$
\begin{mylem}
\label{lem:RandEccAllPartitions}
Given $\mathcal{P}(G)$, the set of all bi-partitions of $G$: 
$$\varepsilon(\mathcal{F}^{ii}_{n-2}) = \sum_{P \in \mathcal{P}(G)}^{i \in V(S)} |\mathcal{T}(S)| |\mathcal{T}(S')| |V(S')|$$
\end{mylem}
\textbf{Proof:} By definition, $\forall F \in \mathcal{F}_{n-2}$, 
$F$ belongs to exactly one of the partitions of $G$. Hence, 
$\mathcal{F}^{ii}_{n-2} = \coprod_{P \in \mathcal{P}}  \varepsilon(\mathcal{F}^{ii}_{n - 2|P})$.
As the RHS is a disjoint union, counting members on both sides we obtain the proof.

\noindent $\square$

\noindent Evidently, combining the results of the two lemmas above, we obtain the proof for 
Theorem \ref{thm:LpPropPartitions}.

\noindent $\square$

\noindent To paraphrase, given a bi-partition $P=(S, S') \in \mathcal{P}(G)$, such that $i \in V(S)$ 
and $j \in V(S')$, Lemma \ref{lem:RandEccPerPartition} yields:  
$\varepsilon(\mathcal{F}^{ii}_{n - 2|P})/\varepsilon(\mathcal{F}^{jj}_{n - 2|P}) = |V(S')|/|V(S)|$. 
Clearly, for a given bi-partition, nodes in the larger of the two components of $P$ have a 
lower number of spanning forests rooted at them than those in the smaller component and vice versa. 
By extension, 
\begin{eqnarray*}
  l^{+}_{ii} - l^{+}_{jj} 
 &\propto& \sum_{P \in \mathcal{P}(G)}^{i \in V(S), j \in V(S')} |\mathcal{T}(S)| |\mathcal{T}(S')| (|V(S')| - |V(S)|)
\label{equ:PropLpForest}
\end{eqnarray*}
can be interpreted as a comparative measure of connectedness of nodes $i$ and $j$. Note that 
for $P \in \mathcal{P}(G)$, the RHS of (\ref{equ:PropLpForest}) is zero when nodes $i$ and $j$ belong 
to the same component of $P$ or if $|V(S)| = |V(S')|$ and positive when 
$i \in V(S), j \in V(S')$ and $|V(S')| > |V(S)|$ or vice versa. 
Therefore,  a node $i$ with higher topological centrality stays connected to 
a greater number of nodes on an average in a disconnected network, than one with lower 
topological centrality and is consequently more immune to random edge failures in the network.

By simple extension, Kirchhoff index represents the average connectedness of all the nodes 
when a failure of a subset of edges partitions the network into two halves, thereby truly reflecting 
overall network robustness. It is easy to demonstrate that of all trees of order $n$, the star has the 
lowest Kirchhoff index and the root of the star has highest $\mathcal{C}^*(i)$ value. Also, amongst all 
graphs of order $n$ with differing volumes, the completely connected graph $K_n$ has the lowest 
Kirchhoff index. 
%
%
\subsection{A Case Study: When the Graph is a Tree}
\label{subSec:CaseStudyTree}
We now study the special case of trees. Recall, a tree $T(V, E)$ of order $n = |V(T)|$ 
is a connected acyclic graph with exactly $n-1 = |E(T)|$ edges. As each of the $n-1$ edges 
$e_{ij} \in E(T)$, upon deletion produces a unique partition $P(S, S') \in \mathcal{P}(T)$, 
we conclude that there are exactly $n - 1$ connected bi-partitions of a tree. Moreover, the 
two sub-graphs $S$ and $S'$ are also trees themselves, such that $|\mathcal{T}(S)| = |\mathcal{T}(S')| = 1$ 
for any partition $P(S, S')$.    
For the nodes of the tree, we then obtain an elegant closed form for topological centrality in 
the following corollary. 
\begin{mycor}
\label{cor:LpPropPartitions}
\begin{equation}
l^+_{ii} = \frac{1}{n^2} \sum_{P \in \mathcal{P}(T)}^{i \in V(S)}  |V(S')|^2 
\end{equation}
\end{mycor}
\textbf{Proof:} The proof follows simply by making the following observations about trees: 
$\varepsilon(\mathcal{F}_{n -1}) = n \cdot 1 = n$. 
Also,  
\begin{equation}
\varepsilon(\mathcal{F}^{ii}_{n-2}) = \sum_{P(S, S') \in \mathcal{P}(G)}^{i \in V(S)}  |V(S')| 
\end{equation}
and
\begin{equation}
\varepsilon(\mathcal{F}_{n-2}) = \sum_{P(S, S') \in \mathcal{P}(G)} |V(S)| |V(S')| 
= \sum_{P(S, S') \in \mathcal{P}(G)} (n - |V(S')|) |V(S')|
\end{equation}  
Thus substituting these values in $(\ref{equ:DenseSRFs})$, we obtain the proof. 

\noindent $\square$

More importantly, in a tree, the shortest path distance $SPD(i, j)$ and the effective resistance 
distance $\Omega_{ij}$ between the node pair $(i, j)$ is exactly the same i.e. 
\begin{equation}
SPD(i, j) = \Omega_{ij} = l^+_{ii} + l^+_{jj} - l^+_{ij} - l^+_{ji}
\end{equation}
The result above is simply due to the fact that a tree is an acyclic graph. It is easy to see that 
\begin{equation}
l^+_{ii} = \sum_{j = 1}^n SPD(i, j) - Tr(\bb L^+) ~~~~ \Rightarrow ~~~~ l^+_{ii} \propto \sum_{j = 1}^n SPD(i, j)
\end{equation}
But the node $i^* \in V(T)$ for which $\sum_{j = 1}^n SPD(i^*, j)$ is the least,  is the so called {\em tree center} 
of $T$. Thus the node with the highest topological centrality, is also the tree center if the graph is a tree; a result
which further knits our centrality measure into the broader body of knowledge (c.f.  \cite{Kirkland97}).   
\

%
%

%
%

\section{Empirical Evaluations}
\label{sec:Empirical}
\begin{figure*}[t]
\centerline{\begin{tabular}{ccc}
\scalebox{0.8}{\includegraphics[width = 90mm]{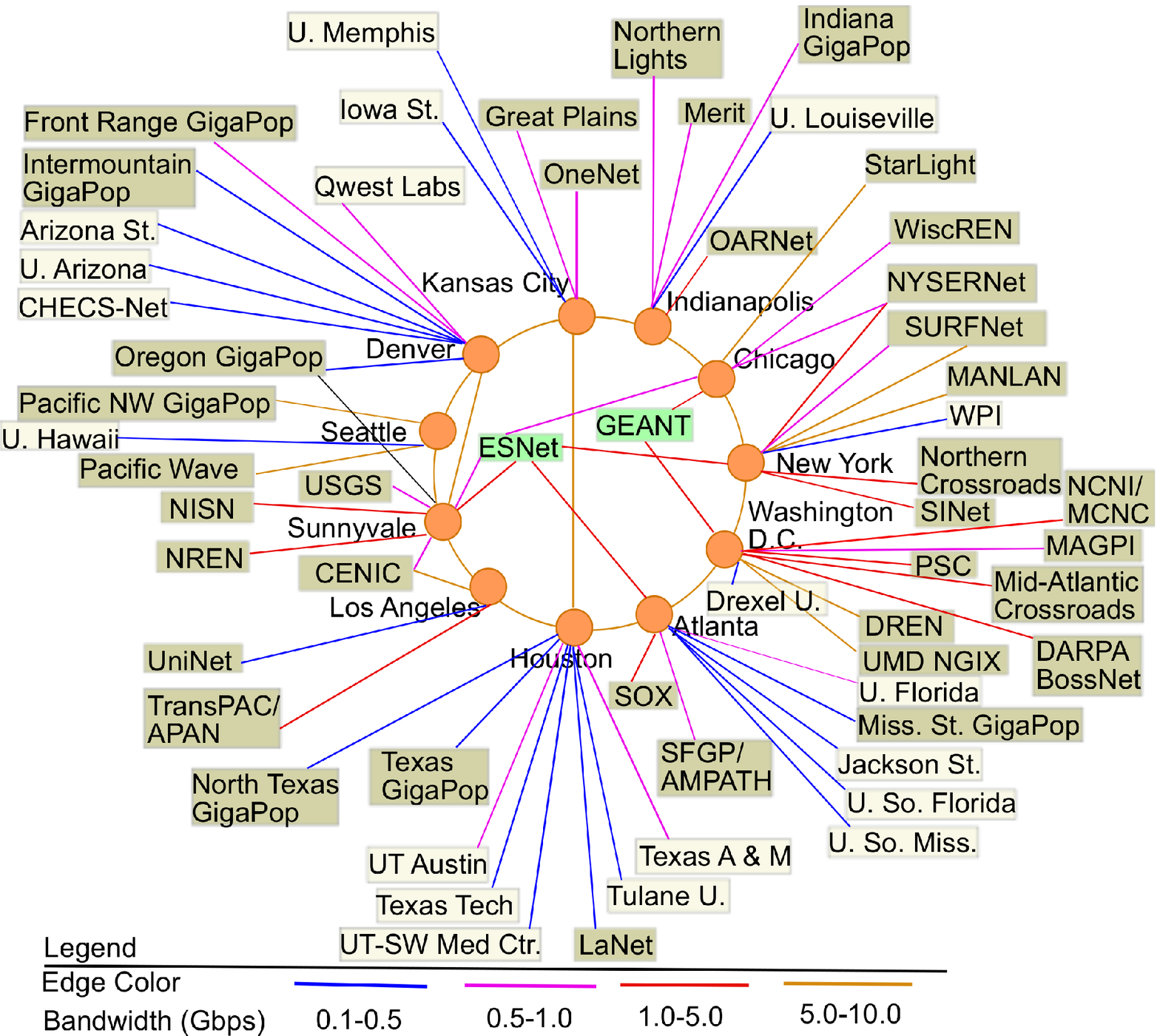}}
&
\scalebox{0.9}{\includegraphics[width = 60mm]{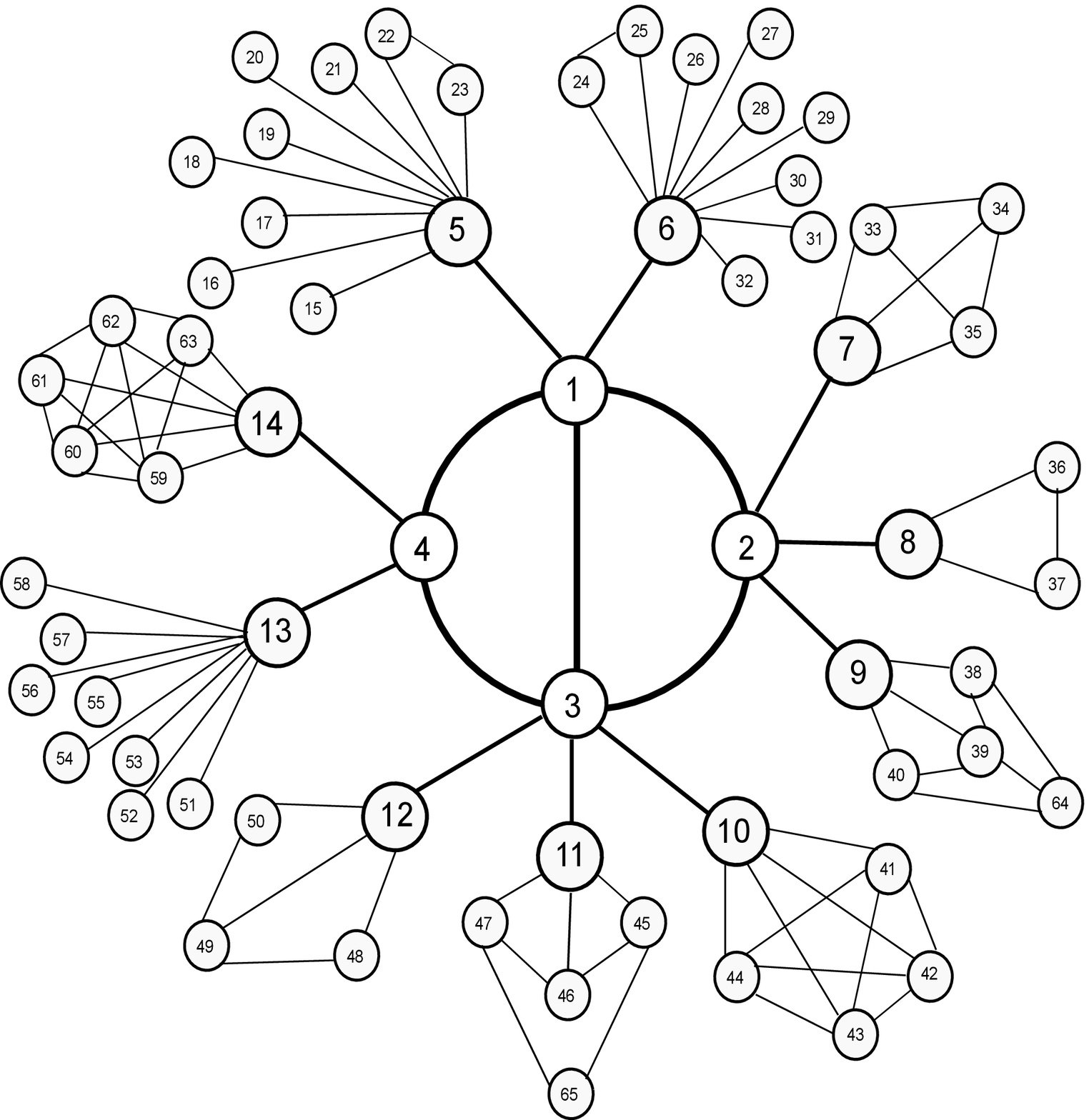}}
&
\scalebox{0.65}{\includegraphics[width = 50mm]{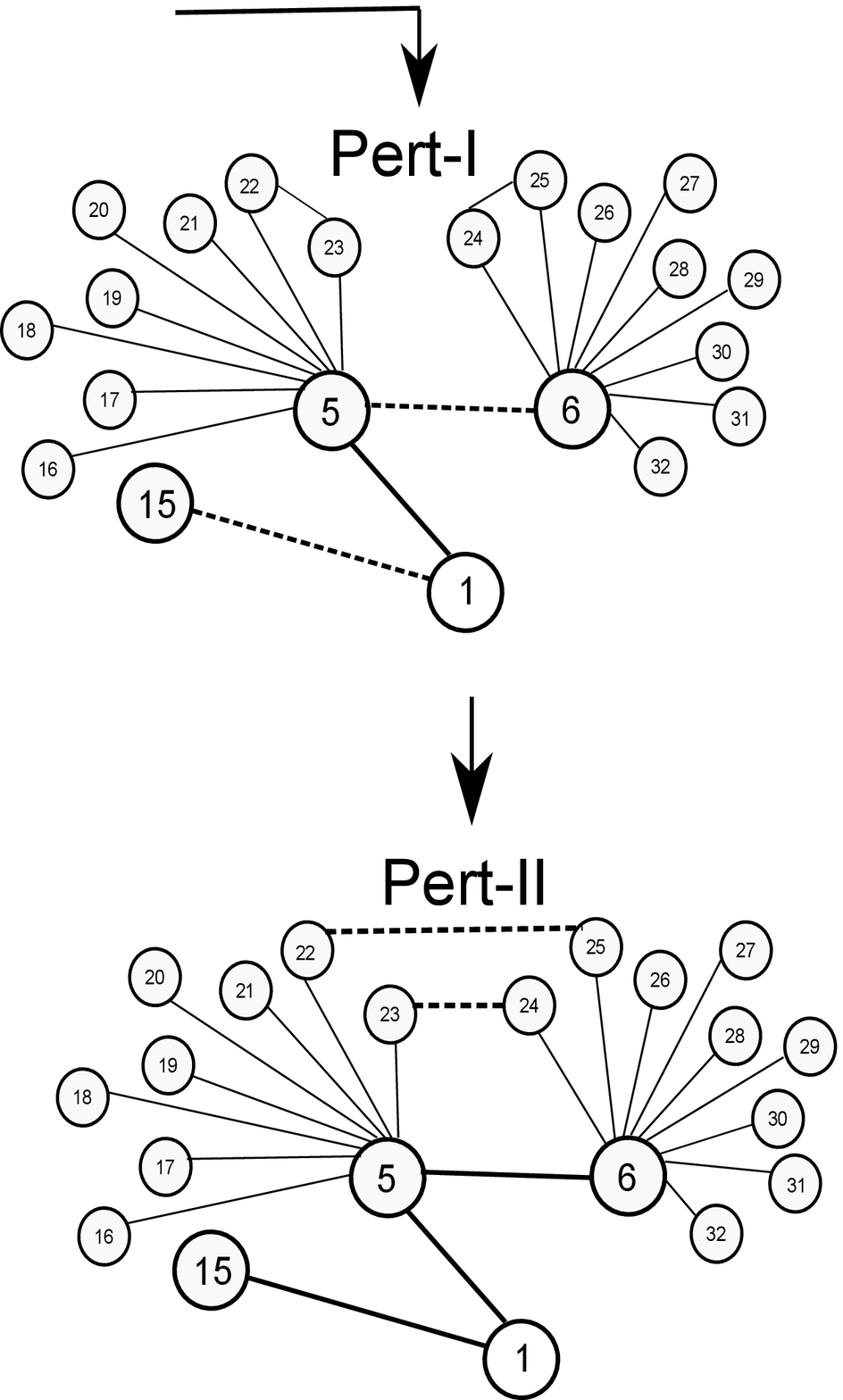}}
\\
(a) Abilene Topology & (b) Simulated topology & (c) Perturbations
\end{tabular}}
\caption{Abilene Network and a simulated topology.}
\label{fig:AbileneTop}
\end{figure*}
\begin{figure*}[h]
\centerline{\begin{tabular}{c}
\scalebox{0.9}{\includegraphics[width = 180mm]{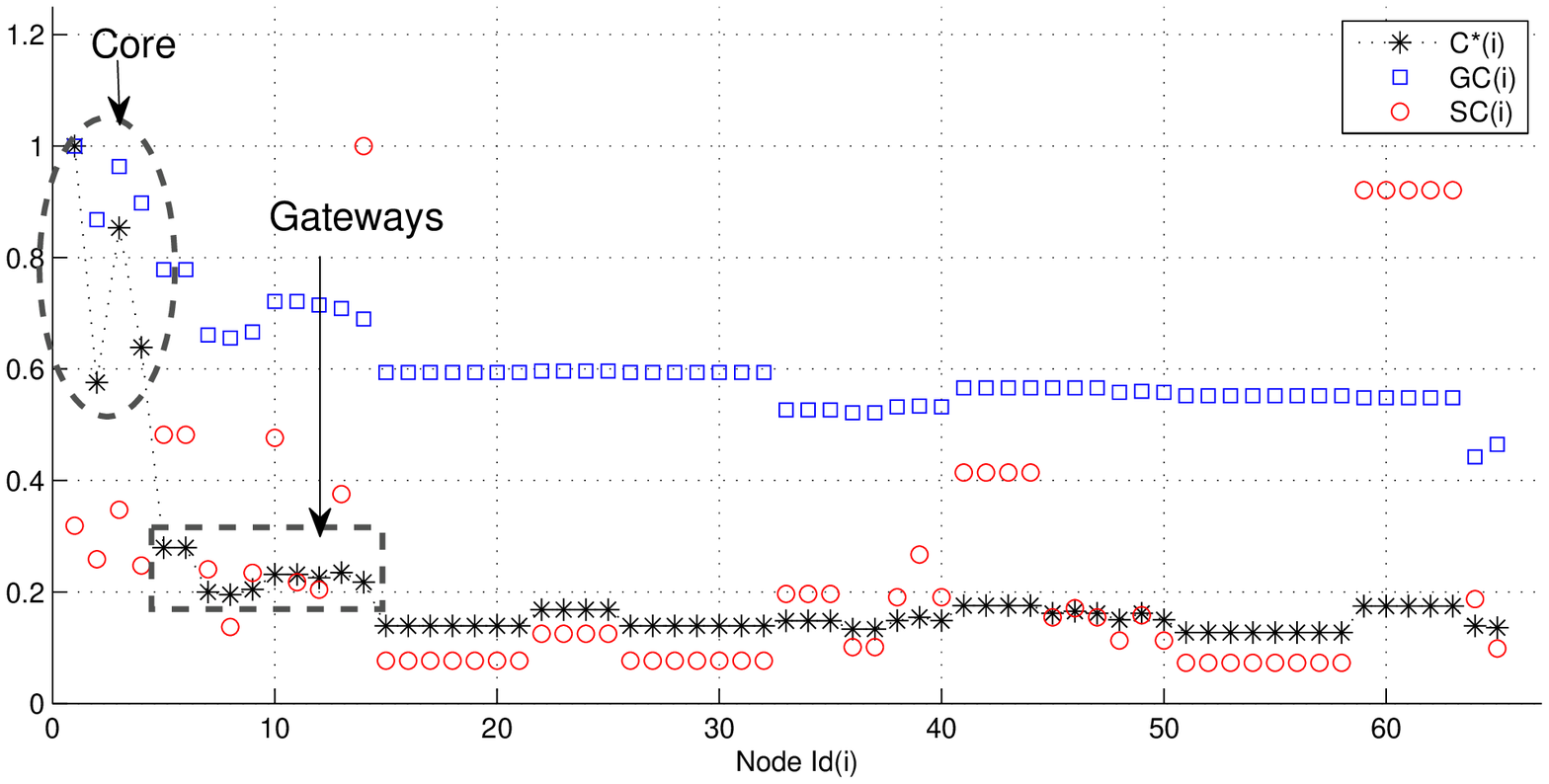}}
\end{tabular}}
\caption{Max-normalized centralities for simulated topology.}
\label{fig:plotTopNets}
\end{figure*}
In this section, we empirically study the properties of topological centrality $(\mathcal{C}^*(i))$ and 
Kirchhoff index (we use $\mathcal{K}^* = \mathcal{K}^{-1}$ henceforth to maintain {\em higher is better}). 
We first show in \textsection\ref{subSec:StructRoles}, how a rank-order of nodes in terms of their 
topological centralities captures their structural roles in the network and then in  
\textsection\ref{subSec:Sensitivity} demonstrate how it, along with Kirchhoff index, is appropriately 
sensitivity to rewirings and local perturbations in the network.
%
%
\subsection{Identifying Structural Roles of Nodes}
\label{subSec:StructRoles}
Consider the router level topology of the Abilene network 
(Fig. \ref{fig:AbileneTop}(a)) \cite{Abilene}. At the core of this topology, 
is a ring of $11$ POP's, spread across mainland US, through which 
several networks interconnect. Clearly, the connectedness of such a 
network is dependent heavily on the low degree nodes on the ring. 
For illustration, we mimic the Abilene topology, with a simulated network 
(Fig. \ref{fig:AbileneTop}(b)) which has a 4-node core $\{v_1,...,v_4\}$ 
that connects $10$ networks through gateway nodes $\{v_5,...,v_{14}\}$ 
(Fig.\ref{fig:AbileneTop}(b)). 

Fig. \ref{fig:plotTopNets} shows the (max-normalized) values of geodesic closeness $(GC)$, 
subgraph centrality $(SC)$ and topological centrality $\mathcal{C}^*$ for the core $\{v_1,...,v_4\}$, 
gateway $\{v_5,...,v_{14}\}$ and other nodes $\{v_{15},...,v_{65}\}$ in topology (Fig.\ref{fig:AbileneTop}(b)). 
Notice that $v_5$ and $v_6$, two of the gateway nodes in the topology, have the highest values of degree
in the network i.e. $(d(v_5)=d(v_6)=10)$ while $v_{14}$ has the highest subgraph centrality $(SC)$. 
In contrast, $\mathcal{C}^*(i)$ ranks the four core nodes higher 
than all the gateway nodes with $v_1$ at the top. The relative peripherality 
of $v_5, v_6$ and  $v_{14}$ as compared to the core nodes requires no 
elaboration. As far as geodesic centrality $(GC)$ is concerned, it ranks all the nodes in the subnetwork 
abstracted by $v_{5}$, namely $v_{15}-v_{23}$, as equals even though $v_{22}$ and $v_{23}$ have 
redundant connectivity to the network through each other and are, ever so slightly, better connected than 
the others - a property reflected in their $\mathcal{C}^*(i)$ rankings. 

\begin{figure*}[ht]
\centerline{\begin{tabular}{cc}
\scalebox{0.7}{\includegraphics[width = 90mm, angle=90]{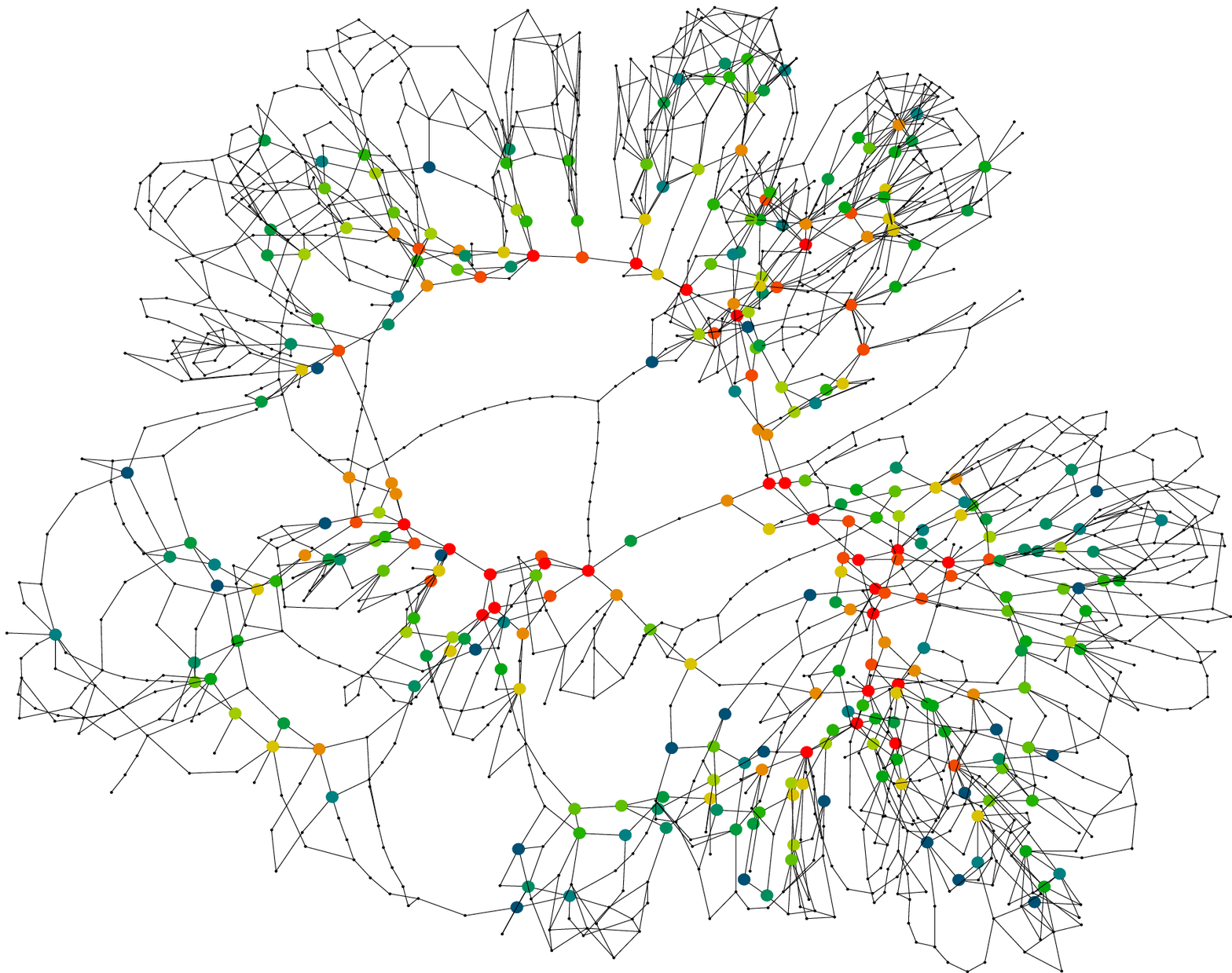}}
&
\scalebox{0.7}{\includegraphics[width = 70mm, angle=90]{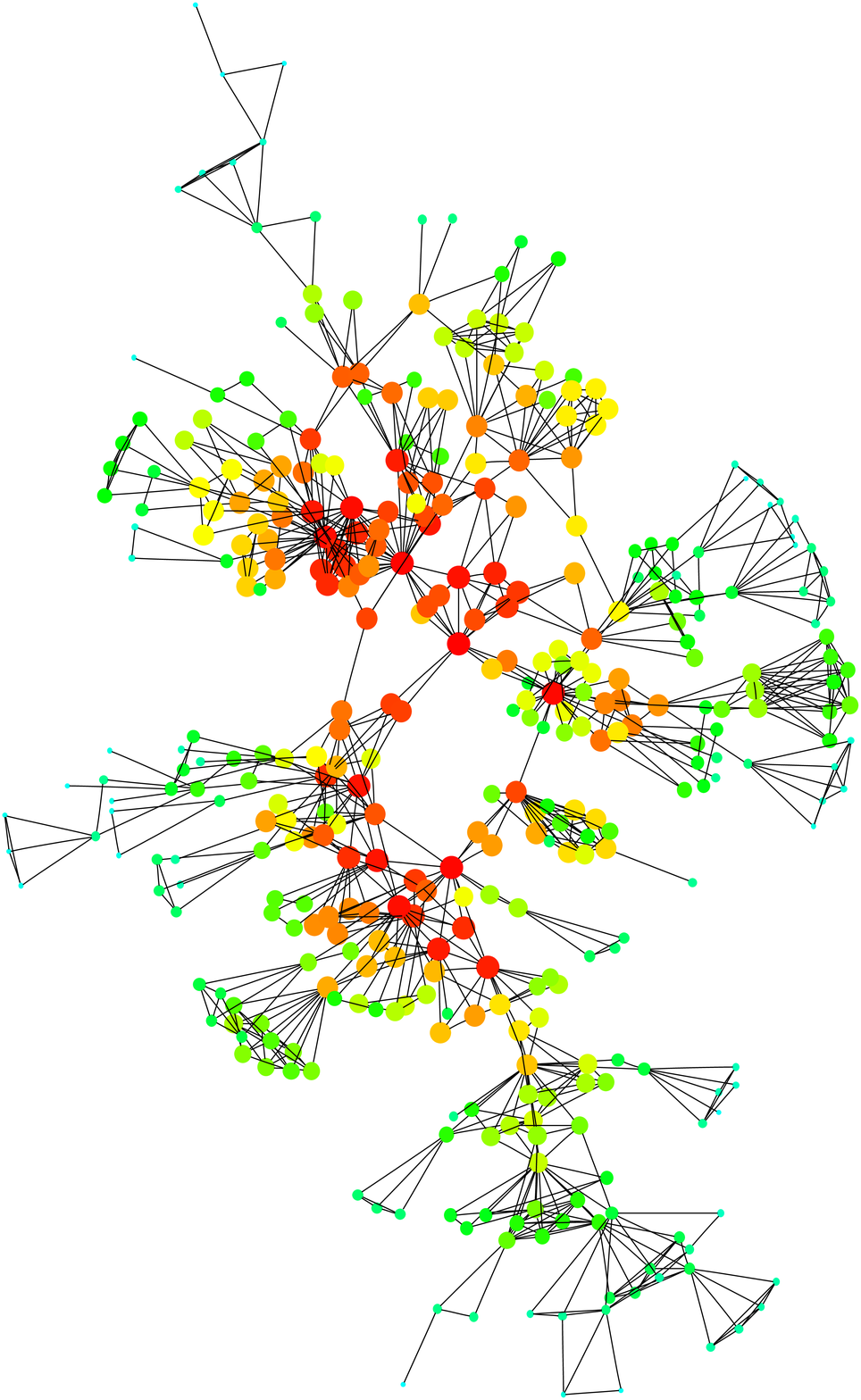}}
\\
(a) The western-states power grid network \cite{Watts98}
&
(b) A network of co-authorships in network sciences \cite{Newman06}
\end{tabular}}
\caption{Real world networks: $Red \rightarrow Turqoise$ in order of decreasing $\mathcal{C}^*(i)$.}
\label{fig:RealWorldNets}
\end{figure*}

We see similar characterization of structural roles of nodes in two real world networks in terms of structural centrality: 
the western states power-grid network \cite{Watts98} and a social network of co-authorships 
\cite{Newman06}, as shown through a color scheme based on $\mathcal{C}^*(i)$ 
values in Fig. \ref{fig:RealWorldNets}. Core-nodes connecting different sub-communities of 
nodes in both these real world networks are recognized effectively by topological centrality 
as being more central ($Red$ end of the spectrum) than several higher degree peripheral nodes.
\
%
%
\subsection{Sensitivity to Local Perturbations}
\label{subSec:Sensitivity}
\begin{figure*}[ht]
\centerline{\begin{tabular}{cc}
\scalebox{0.8}{\includegraphics[width = 110mm]{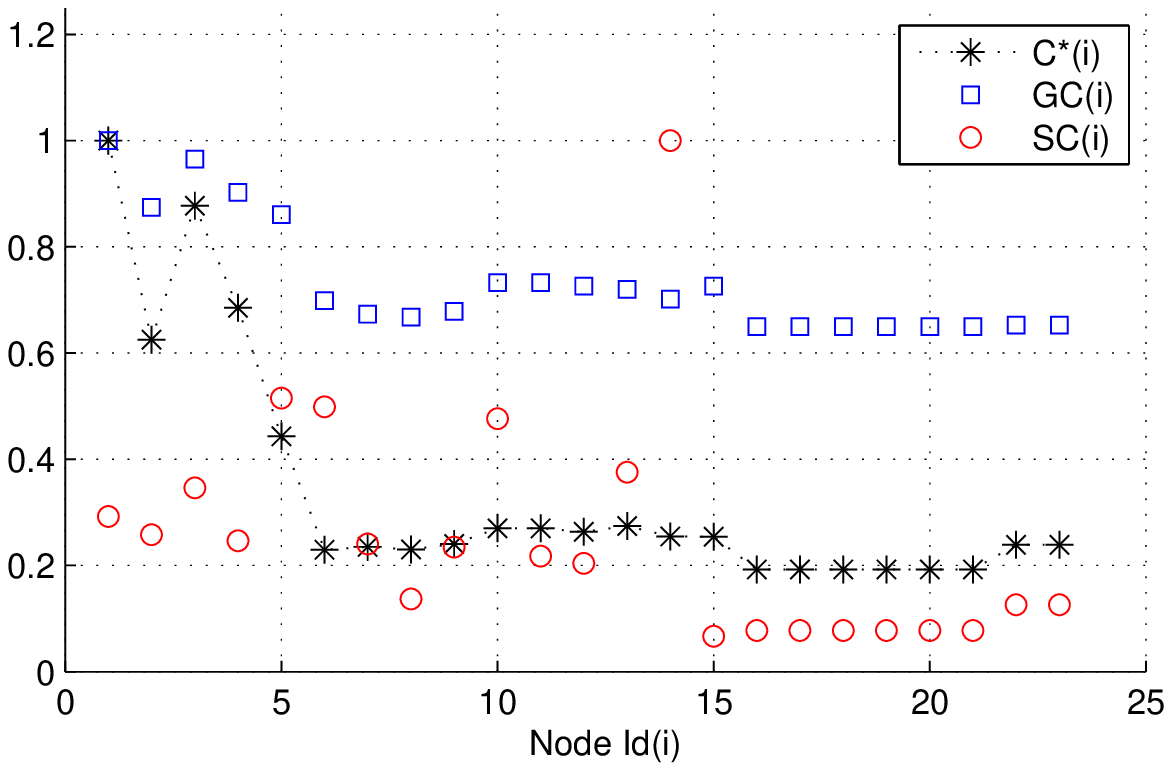}}
&
\scalebox{0.8}{\includegraphics[width = 110mm]{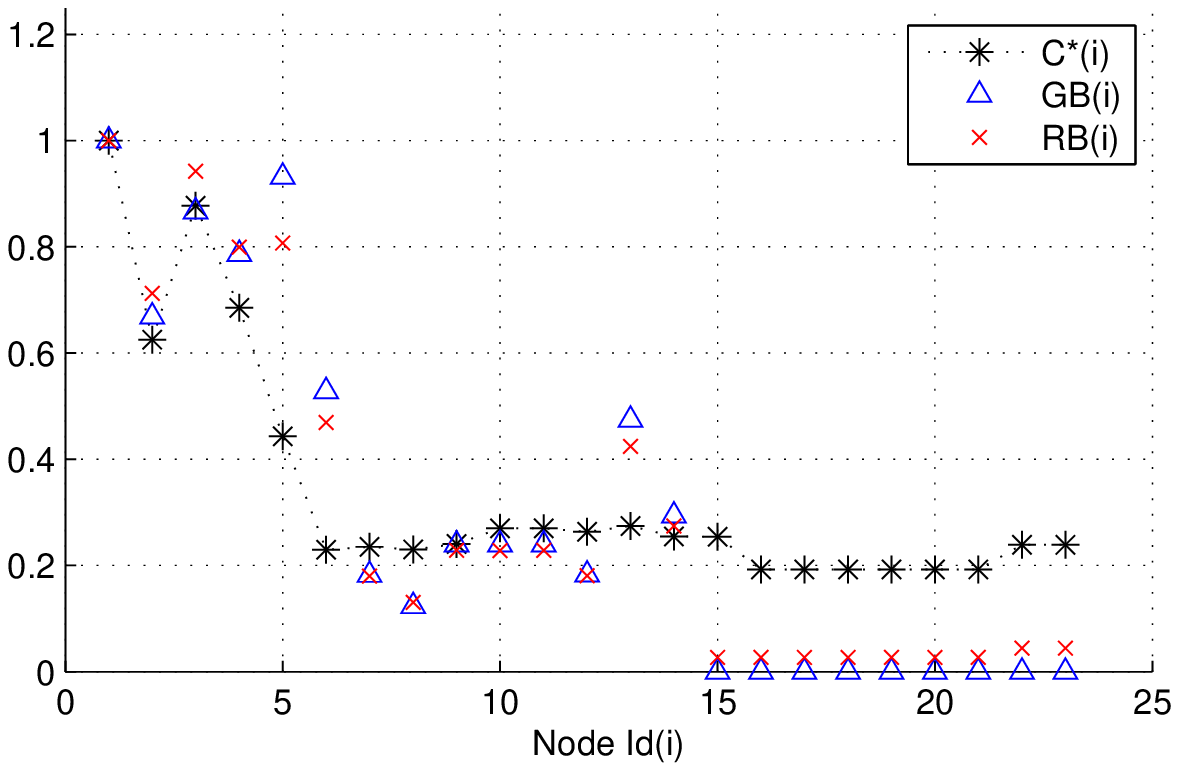}}
\\
(a) Vis-\`{a}-vis centralities & (b) Vis-\`{a}-vis betweennesses
\end{tabular}}
\caption{PERT-I: Max-normalized values of centralities and betweennesses for core, gateway and some other nodes.}
\label{fig:plotTopNets_HighStat}
\end{figure*}
An important property of centrality measures is their sensitivity to perturbations in 
network structure. 
Traditionally, structural properties in real world networks have been equated to 
average statistical properties like power-law/scale-free degree distributions and 
rich club connectivity \cite{Barabasi00, Faloutsos99, Barabasi01}.  
However, the same degree sequence $D = \{d(1)\geq d(2)\geq ...\geq d(n)\}$, 
can result in graphs of significantly varying topologies. 
Let $\mathcal{G}(D)$ be the set of all connected graphs with scaling 
sequence $D$. The generalized Randi\'{c} index $R_1(G)$ 
\cite{BollobasErdos98, Randic75}:
\begin{equation}
R_1(G) = \sum_{e_{ij} \in E(G)} d(i) d(j) \label{equ:S-Fact}
\end{equation}
where $G \in \mathcal{G}(D)$, is  considered to be a measure of overall 
connectedness of $G$. Higher $R_1(G)$ suggests that nodes of higher degrees 
connect with each other with high probability thereby displaying the so-called 
{\em rich club connectivity} (RCC) in $G$ \cite{LiAldersonWillingerDoyle04}. 
Similarly, the average of each centrality/betweenness index 
($GC, SC, GB, RB$ averaged over the set of nodes), is in itself a global structural 
descriptor for the graph $G$ \cite{Estrada05}. 
We now examine the sensitivity of each index with respect to local perturbations 
in the subnetwork abstracted by the core node $v_1$ and its two gateway neighbors 
$v_5$ and $v_6$.   

First, we rewire edges $e_{15,5}$ and $e_{6,1}$ to $e_{15,1}$ and $e_{6,5}$ 
respectively (PERT-I Fig. \ref{fig:AbileneTop}(c)). PERT-I is a degree  
preserving rewiring which only alters local connectivities i.e. neither individual node degrees 
nor average node degree changes.
Fig. \ref{fig:plotTopNets_HighStat}(a) and (b) 
respectively show the altered values of centralities $(\mathcal{C}^*, GC, SC)$ and 
betweennesses, geodesic and random-walk i.e. $(GB, RB)$, after PERT-I.
Note, after  PERT-I,  $v_{15}$ is directly connected to $v_1$ which makes $\mathcal{C}^*(v_{15})$ 
comparable to other gateway nodes while $SC(v_{15}), GB(v_{15}), RB(v_{15})$ seem to be entirely 
unaffected.
Moreover, PERT-I also results in $v_6$ losing its direct link to the core, reflected in the 
decrease in $\mathcal{C}^*(v_6)$ and a corresponding increase in $\mathcal{C}^*(v_5)$. 
$\mathcal{C}^*(i)$, however, still ranks the core nodes higher than $v_5$ 
(whereas $SC, GB, RB$ do not) because PERT-I being a local perturbation should not affect 
nodes outside the sub-network --- 
$v_1$ continues to abstract the same sub-networks from the rest of the 
topology. We, therefore, observe that $\mathcal{C}^*(i)$ is appropriately 
sensitive to the changes in connectedness of nodes in the event of local 
perturbations.  But what about the network on a whole?
\begin{table}[t]
\caption{Sensitivity to local perturbations, 
$\overline{X} = 1/n \sum_i^n X(i)$: Avg. node centrality for a network.}
\label{Tab:Sensitivity_To_Pert}
\begin{center}
\small
\begin{tabular}{||l|l||c|c||}
\hline
$\#$ & Structural Descriptor & PERT-I  & PERT-II \\
\hline\hline
1 & $\mathcal{K}^*(G)$ (or $\overline{\mathcal{C}^*}$) & $\downarrow$ & $\uparrow$ \\
\hline
2 & $R_1(G)$ & $\uparrow$ & $\leftrightarrow$ \\
\hline
3 & $\overline{GC}$ & $\downarrow$ & $\uparrow$ \\
\hline
4 & $\overline{SC}$, $\overline{GB}$  & $\uparrow$ & $\downarrow$ \\
\hline
5 & $\overline{RB}$ & $\uparrow$ & $\uparrow$ \\
\hline
\end{tabular}
\end{center}
\end{table}
Let $G$ and $G_1$ be the topologies before and after PERT-I. $G_1$ is less 
well connected overall than $G$  as the failure of $e_{5,1}$ in $G_1$ 
disconnects $19$ nodes from the rest of the network as compared to $10$ nodes 
in $G$. However,
$$\Delta R_1(G \rightarrow G_1) = \frac{R_1(G_1) - R_1(G)} {R_1(G)} = 0.029$$ 
as the two highest degree nodes ($v_5$ and $v_6$) are directly connected in 
$G_1$ (see TABLE \ref{Tab:Sensitivity_To_Pert} for the sensitivity of other centrality based 
global structural descriptors).
In contrast, $\Delta \mathcal{K}^*(G \rightarrow G_1) = -0.045$, which rightly  
reflects the depreciation in overall connectedness after PERT-I 
(recall  $\mathcal{K}^*(G) =  \mathcal{K}^{-1}(G$). 
Table \ref{Tab:Sensitivity_To_Pert} shows the changes in the average of all centrality and 
betweenness indices post PERT-I. 

A subsequent degree preserving perturbation PERT-II of $G_1$, 
rewiring $e_{22,23}$ and $e_{24,25}$ to $e_{22,25}$ and $e_{23,24}$,
to obtain $G_2$, creates two cycles in $G_2$ that safeguard against the failure 
of edge $e_{5,6}$. This significantly improves local connectivities in the sub-network.  
However, $\Delta R_1(G_1 \rightarrow G_2) = 0$ (and average $SC$ decreases) 
while $\Delta \mathcal{K}^*(G_1 \rightarrow G_2) = 0.036$ which once again shows 
the efficacy of Kirchhoff index as a measure of global connectedness of networks.
\

%
%

%
%

\section{A Word on Computational Complexity}
\label{sec:Complexity}
We now discuss the practical aspects of computing the topological centrality measure for the set of nodes in 
the graph representing the complex network. It suffices to compute the pseudo-inverse of the matrix $\bb L$, 
the Laplacian, to obtain the matrix $\bb L^+$. 
The most common method for computing the pseudo-inverse of a matrix mathematically, is to use the 
singular value decomposition (SVD). Indeed, mathematical software such as Matlab, come 
equipped with subroutines, such as {\em pinv} (for pseudo-inverse), which make use of the SVD 
factorization. It is common knowledge that the computational complexity for the SVD method is, in terms 
of worst case complexities, $O(n^3)$ where $n$ is the number of rows/columns of the matrix. Thus the base 
worst case complexity for computing the topological centrality measure is indeed $O(n^3)$, where $n$ 
is the order of the graph. This worst case complexity is at par with the competitive centrality measures, 
like geodesic centrality $(GC)$ and betweenness $(GB)$ (based on shortest paths) as well as subgraph 
centrality $(SC)$; and better than the random-walk betweenness $(RB)$ (see Table \ref{tab:Taxonomy}).     

However, given that graphs (or networks abstracted as graphs), are topological objects, and the nature 
of the eigen spaces of the matrices $\bb L$ and $\bb L^+$ (as discussed in \textsection\ref{sec:Geometry}), 
we can be clever from the computational point of view.  Although a full discussion of these computational 
improvisations is out of the scope of the current work, we provide some useful pointers for the interested 
reader. 
Exploiting the fact that $\lambda_n = 0$, the smallest eigen value of $\bb L$, is unique if the network is 
connected, it has been shown in \cite{Xiao03} that $\bb L^+$ can be computed by perturbing the matrix 
$\bb L$ (by adding a constant $1/n$ to every element) which yields an {\em invertible} full rank matrix. This 
method, though considerably faster in practice on Matlab, still leaves the worst case complexity at $O(n^3)$.  

In \cite{Fouss07}, we find a way of approximating $\bb L^+$ by using fast converging Monte-Carlo 
algorithms. Such parallel algorithms exploit the sparsity of real world networks which in turn makes 
the Laplacian $\bb L$ a sparse matrix (even though $\bb L^+$ is always full). 
However, we observe that from the point of view of computing topological 
centrality alone $(\mathcal{C}^*(i))$, all we need is the diagonal of $\bb L^+$. 
It is known that subsets of inverse for a sparse matrix can be computed to a given pattern 
(selective elements), using parallel and multi-frontal 
approaches (c.f. \cite{Campbell95} and the references therein). 

Finally, another interesting result has recently been proposed in \cite{Luxburg10} that shows that as 
the density of edges in the network increases, the hitting time from node $i$ to $j$ can be 
well approximated as $H_{ij} \approx Vol(G) d(i)^{-1}$. By extension, the commute time becomes 
$C_{ij}  \approx Vol(G) (d(i)^{-1} + d(j)^{-1})$.  
So for dense graphs, we can compute topological centrality for nodes using the node degree distribution 
alone (see \textsection\ref{sec:StructCentAndRandomWalks}).   
Most importantly, this result implies that topological centrality, which is a measure of the overall position and 
connectedness of a node, is determined entirely by its local connectedness determined by its degree, 
a remarkable result indeed. 
\

%
%

%
%

\section{Conclusion and Future Work}
\label{sec:Conclusion}
In this work, we presented a geometric perspective on robustness in complex  
networks in terms of the Moore-Penrose pseudo-inverse of the graph Laplacian. 
We proposed topological centrality $(\mathcal{C}^*(i))$ and Kirchhoff 
index $(\mathcal{K}(G))$ that respectively reflect the length of the position vector 
for a node and the overall volume of the graph embedding and therefore are 
suitable geometric measures of robustness of individual nodes and the overall network. 
Additionally, we provided interpretations for these indices 
in terms of the overhead incurred in random detours through a node in question 
as well as in terms of the recurrence probabilities and voltage distribution in the 
EEN corresponding to the network. Both indices reflect the global connectedness 
properties of individual nodes and the network on a whole, particularly in the event 
of multiple edge failures that may render the network disconnected. 
Through numerical analysis on simulated and real world networks, we demonstrated 
that $\mathcal{C}^*(i)$ captures structural roles played by nodes in networks and, 
along with Kirchhoff index, is suitably sensitive to perturbations/rewirings in 
the network. In terms of computational complexity, topological centrality 
compares well with other geodesic and all-paths based indices in literature 
(see Table \ref{tab:Taxonomy}) and performs better than random-walk betweenness 
in the asymptotic case. 
In future, we aim at investigating similar metrics for the case of 
strongly connected weighted directed graphs to further generalize our work, a preliminary 
attempt towards which has already begun in the form of the results in \cite{Boley11}.   
\begin{table}
\caption{Taxonomy and computational complexities of centrality measures (for all nodes).}
\label{tab:Taxonomy}
\centerline{
\begin{tabular}
{|c|l|c|c|}
\hline
\# & Measure & Paths covered & Complexity \\
\hline \hline
1. & Degree & - & $O(m)$\\
\hline
2. & $GC$,  $GB$ & Geodesic paths & $O(n^3)$\\
\hline\hline
3. & $\mathcal{C}^*$ & All paths & $O(n^3)$\\
\hline
4. & $SC$ & All paths & $O(n^3)$\\
\hline
5. & $RB$  & All paths & $O(m + n) n^2$\\ 
\hline
6. & $FB$ & All paths & $O(m^2 n)$ \\
\hline
\end{tabular}}
\end{table}
\

%
%

%
%

\section{Acknowledgment}
\label{sec:Acknowledgment}
We express our sincere gratitude towards the anonymous reviewers whose suggestions 
helped improve the quality of this work. 
This research was supported in part by 
DTRA grant HDTRA1-09-1-0050 and NSF grants CNS-0905037, CNS-1017647 and CNS-1017092.
\

%
%

%
%

\bibliography{allrefs}
\bibliographystyle{abbrv}

%
%

%
%

%
%

%
%

\end{document}